\documentclass[12pt,a4paper]{article}
\usepackage[latin1]{inputenc}
\usepackage{amsmath}
\usepackage{amsfonts}
\usepackage{amssymb}
\usepackage{graphicx}
\usepackage{wrapfig}
\usepackage[a4paper,margin=1.0in]{geometry}

\begin{document}

\title{Positron Source Simulations for ILC 1 TeV Upgrade\thanks{ Work supported by the German Federal Ministry of Education and
Research, Joint Research Project R\&D Accelerator ``Spin Optimization'',
contract number 19XL7IC4}}

\author{Andriy Ushakov\thanks{ andriy.ushakov@uni-hamburg.de}, Gudrid Moortgat-Pick, \textit{University of Hamburg, Germany}\\
Sabine Riemann, \textit{DESY, Zeuthen, Germany}\\
Wanming Liu, Wei Gai, \textit{Argonne National Laboratory, USA}}

\date{}

\maketitle{}
\vspace{-65mm}
\hfill DESY-13-003
\vspace{65mm}

\begin{abstract}
The generation and capture of polarized positrons at a source with a superconducting helical undulator having 4.3\,cm period and 500\,GeV electron drive beam have been simulated. The positron polarization has been calculated for the different undulator $K$ values (up to $K = 2.5$). Without applying a photon collimator, the maximal polarization of positrons is about 25\% for 231 meters active magnet length of undulator with $K = 0.7$. Using an undulator with $K = 2.5$ and a collimator with an aperture radius of 0.9\,mm results in increase of positron polarization to 54\%.
The energy deposition, temperature rise and stress induced by high intense photon beam in the rotated titanium-alloy target have been estimated. The maximal thermal stress in the target is about 224\,MPa for the source with photon collimation to achieve a  positron polarization of 54\%.
\end{abstract}

\section{Introduction}

The current design for the future International Linear Collider (ILC) includes a positron source based on a superconducting helical undulator which is placed at the end of main linear accelerator. Due to the helical undulator the generated photons are circularly polarized and created positrons are longitudinally polarized. The degree of polarization is determined by the undulator and electron beam parameters.

A prototype of a helical undulator for the ILC positron source has been developed and tested at Daresbury \cite{ILCundulator}. According to the ILC requirements \cite{ILC}, the yield of the source should have 50\% safety margin in a wide energy range of drive beam energy (between 100\,GeV and 250\,GeV). That means the positron yield at the injection point into the Dumping Ring (DR) has to be 1.5 positrons per electron going through the undulator. A center-of-mass energy of 1\,TeV is considered as upgrade option.

\begin{figure}[htb]
  \centering
  \includegraphics*[width=78mm]{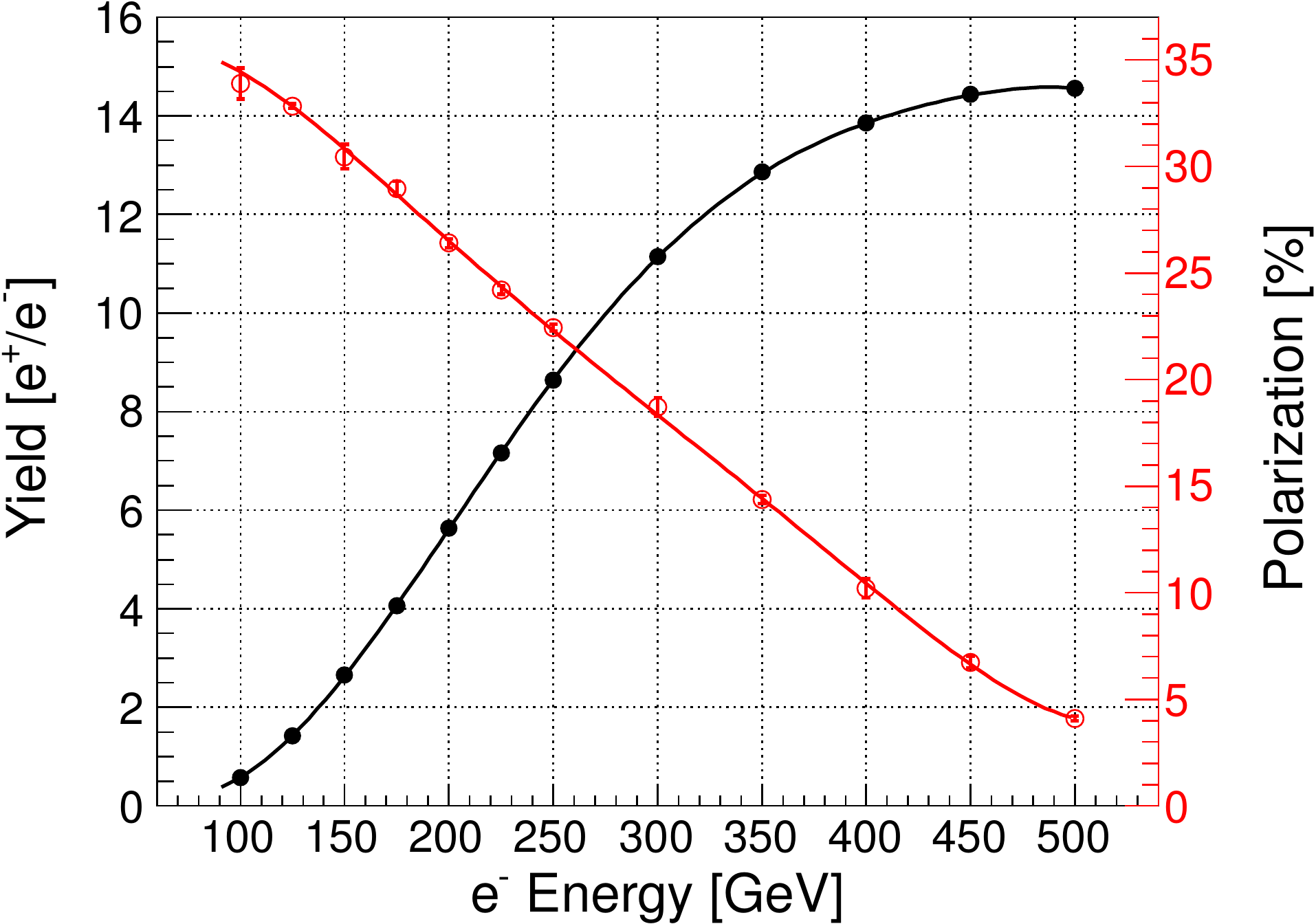}$~~~~$
  \includegraphics*[width=75mm]{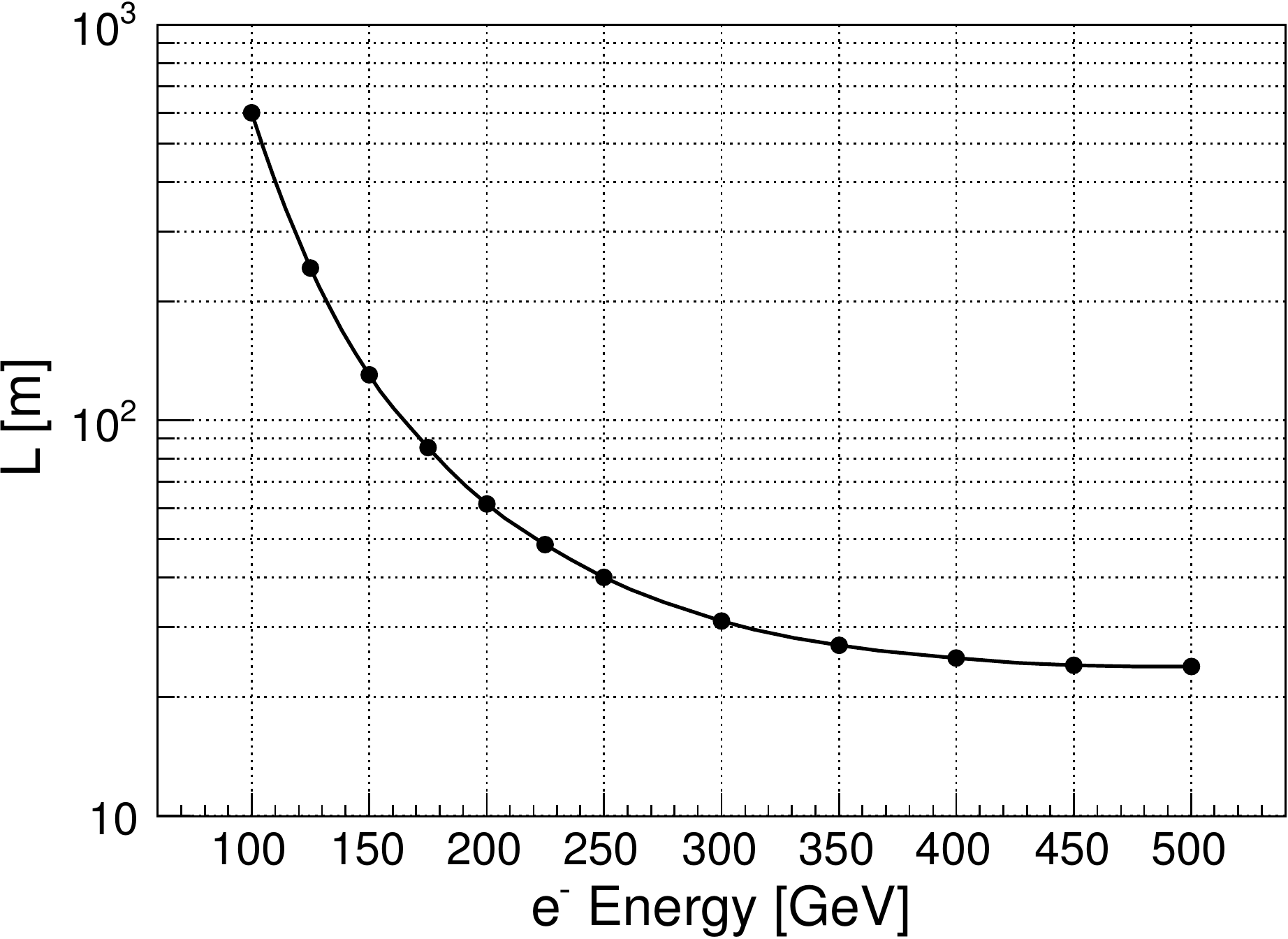}
  \caption{Positron yield and polarization of the positron source with 231\,m RDR undulator and 3.2\,Tesla peak field of pulsed flux concentrator (left) and undulator length required for a yield of 1.5\,e$^+$/e$^-$ (right).}
  \label{fYieldPolUlength-vs-E}
\end{figure}

Figure\,\ref{fYieldPolUlength-vs-E} shows the positron yield depending on the electron beam (drive) energy for a source with a fixed undulator length of $L = 231\,$m, an undulator period $\lambda = 11.5~$mm and $K = 0.92$. The source with these undulator parameters (RDR undulator) can generate much more positrons than required, therefore, there are two ways to keep the yield at 1.5 e$^+$/e$^-$, either to reduce the undulator length by switching-off unnecessary modules (see the right plot in Fig.\,\ref{fYieldPolUlength-vs-E}) or to reduce the magnetic field of the undulator. 

\begin{figure}[htb]
  \centering
  \includegraphics*[width=75mm]{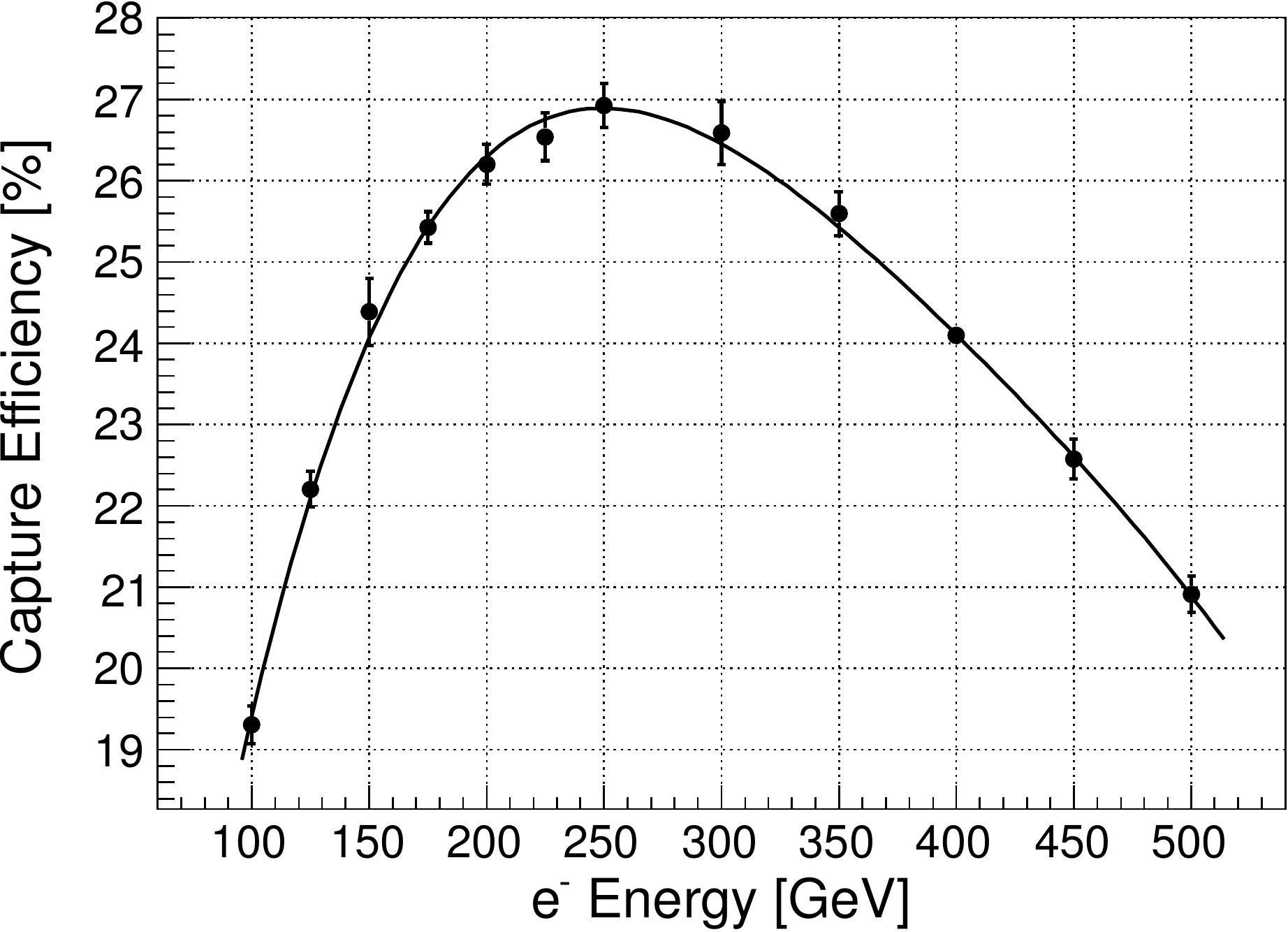}
  \caption{Capture efficiency of the source with RDR undulator and 3.2 Tesla peak field of pulsed flux concentrator.}
  \label{fCErdr-vs-E}
\end{figure}

The source parameters, especially the target thickness and positron capture optics, have been optimized for getting maximal positron yield for a 250\,GeV drive beam:
the titanium alloy (Ti6Al4V) target has a thickness of 0.4 radiation length; the 12 cm long pulsed flux concentrator has 3.2 Tesla maximal field. These parameters have been kept in all our calculations presented in this report. The simulations have been performed by a Geant4-based tool that was specially developed for Polarized Positron Source Simulations (PPS-Sim) \cite{pps-sim}. Figure\,\ref{fCErdr-vs-E} shows the capture efficiency of the source with RDR undulator. The capture efficiency is the ratio of the number of positrons at the end of the source (the positron beam has to fit DR acceptance) to the number of positrons after the target. The maximum of capture efficiency is about 27\% at 250\,GeV. For a 500\,GeV e$^-$ beam the capture efficiency is falling down to 21\%.

Figure\,\ref{fYieldPolUlength-vs-E} shows the dependence of polarization on the $e^-$ energy (see the red curve on the left plot). The higher $e^-$ energy results in lower polarization. 5\% polarization at 500\,GeV is too low to get any advantages for physics of using such low polarized positron beams. 

The opening angle of undulator radiation (as well as the radius of photon spot size on target) is inversely proportional to the electron energy. For instance, doubled e$^-$ energy results in four times higher energy deposition density in a stationary target. With lower $K$ values  higher polarization can be achieved.
However, lowering the undulator $B$ field will additionally reduce the photon spot size. Therefore, for the 1\,TeV upgrade of the ILC, another undulator with higher period has been proposed in \cite{Wanming1TeV}. 

In this report, the maximum achievable polarization of a positron source using a 4.3\,cm period undulator and configurations with and without photon collimator has been estimated. The energy deposition and thermal stress in the target has been simulated.

\section{Yield and Polarization of a Source with 4.3\,cm Period Undulator }

The e$^+$ yield and polarization of a source at 500\,GeV e$^-$ and with different undulator periods have been estimated earlier (see Ref.\,\cite{Wanming1TeV}). In this report, the dependence on the undulator $K$ value  will be analyzed for the selected 4.3\,cm undulator period.

\begin{figure}[htb]
  \begin{center}
  \includegraphics*[width=75mm]{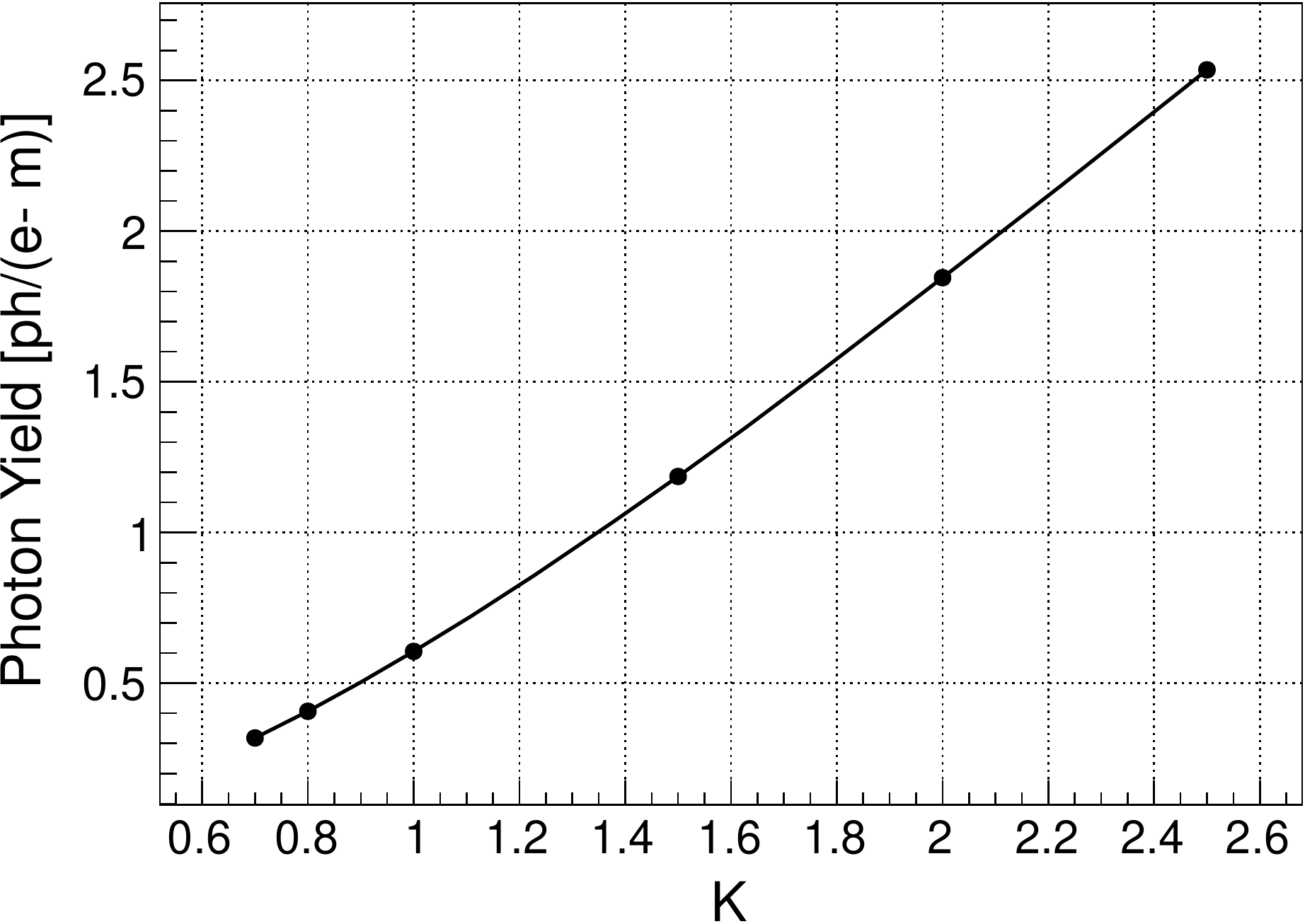}$~~~~~~$
  \includegraphics*[width=75mm]{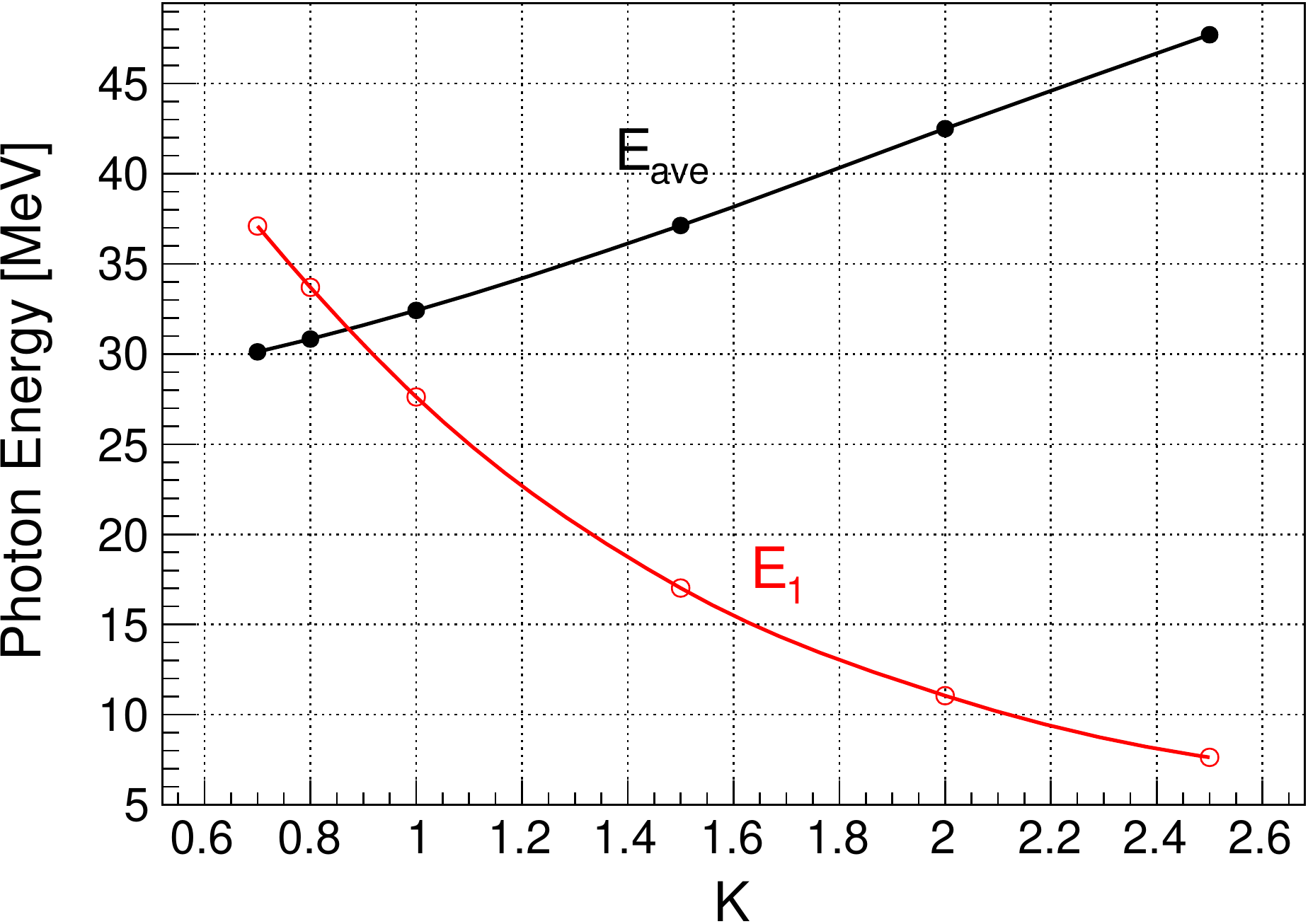}
  \caption{Photon yield (left) and energy of photons (right) vs undulator $K$-value. $E_{\mathrm{ave}}$ is the average photon energy and $E_1$ is the energy cutoff of the $1^{\mathrm{st}}$ harmonic.}
  \label{fPhYieldEn-vs-K}
  \end{center}
\end{figure}

In PPS-Sim, the implementation of undulator radiation is based on Kincaid's model~\cite{Kincaid}. The efficiency of photon generation in the undulator having different $K$ values is shown in Fig.\,\ref{fPhYieldEn-vs-K} (left plot). The photon yield has been normalized per electron and meter of undulator. The photon energy cut-off of the 1st harmonic and the average photon energy are also shown in Fig.\,\ref{fPhYieldEn-vs-K} (right plot).

An undulator with higher $K$ value yields lower energy of the fundamental harmonic but larger contribution of higher harmonics. As result the average energy over the whole photon spectrum is growing with increasing field of undulator.

Both tendencies (yield and average energy versus $K$) indicate that an electron beam passing an undulator with higher $K$ generates a positron beam with higher current.

\begin{figure}[htb]
  \centering
  \includegraphics*[width=75mm]{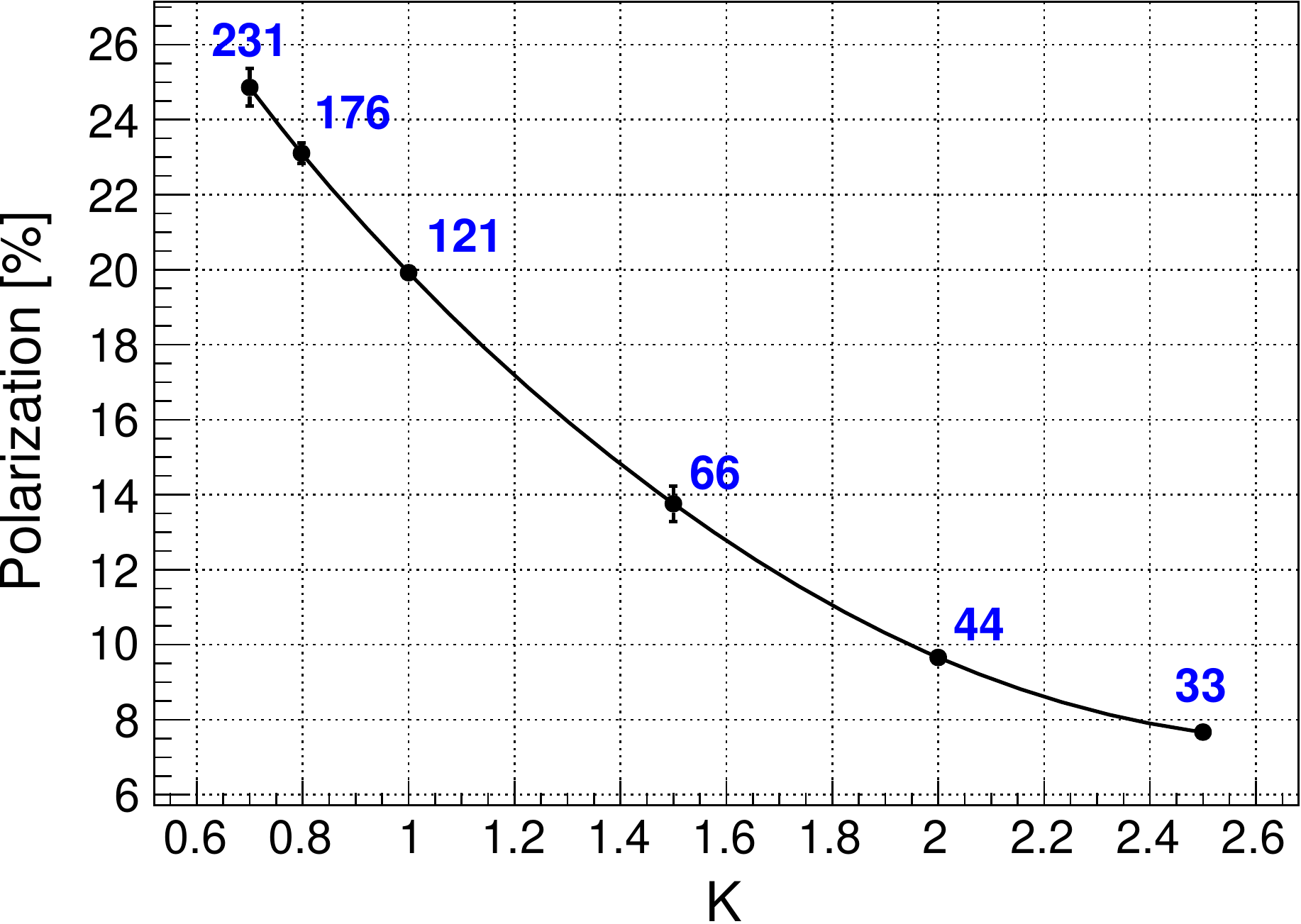}
  \caption{Positron polarization vs $K$ for a source without collimator and with FC having 3.2 T peak field on beam axis. The blue numbers indicate the required undulator length.}
  \label{fPol-vs-K-woC}
\end{figure}

The impact of the undulator field on the $e^+$ polarization is shown in Fig.\,\ref{fPol-vs-K-woC}. In these simulations we suppose that the magnet length of an undulator cryomodule is 11 meters and the drift space between the end of the undulator and the target is 412 meters. If not all modules are necessary upstream modules are switched off. 

\begin{figure}[htb]
  \centering
  \includegraphics*[width=75mm]{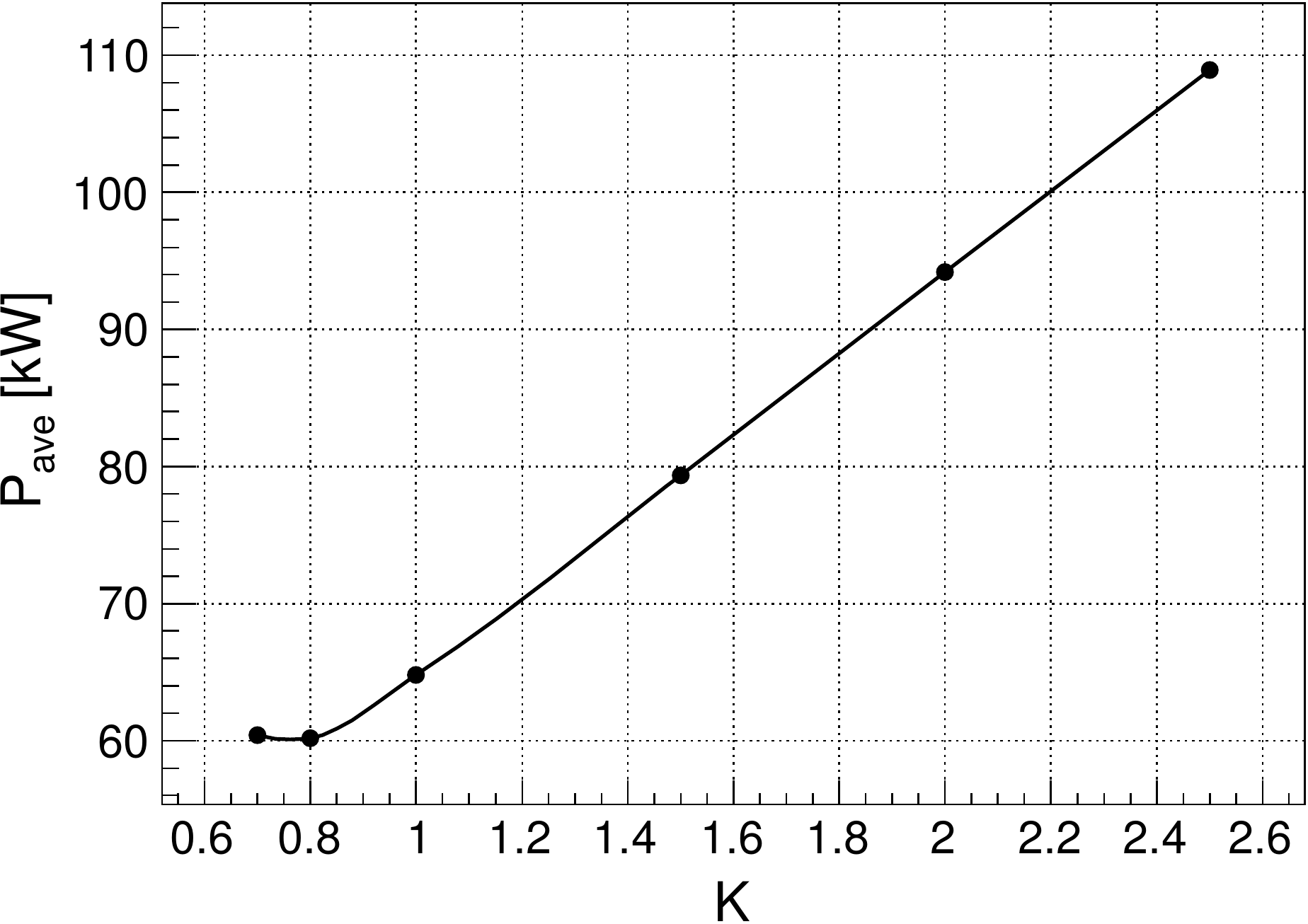}$~~~~~~$
  \includegraphics*[width=75mm]{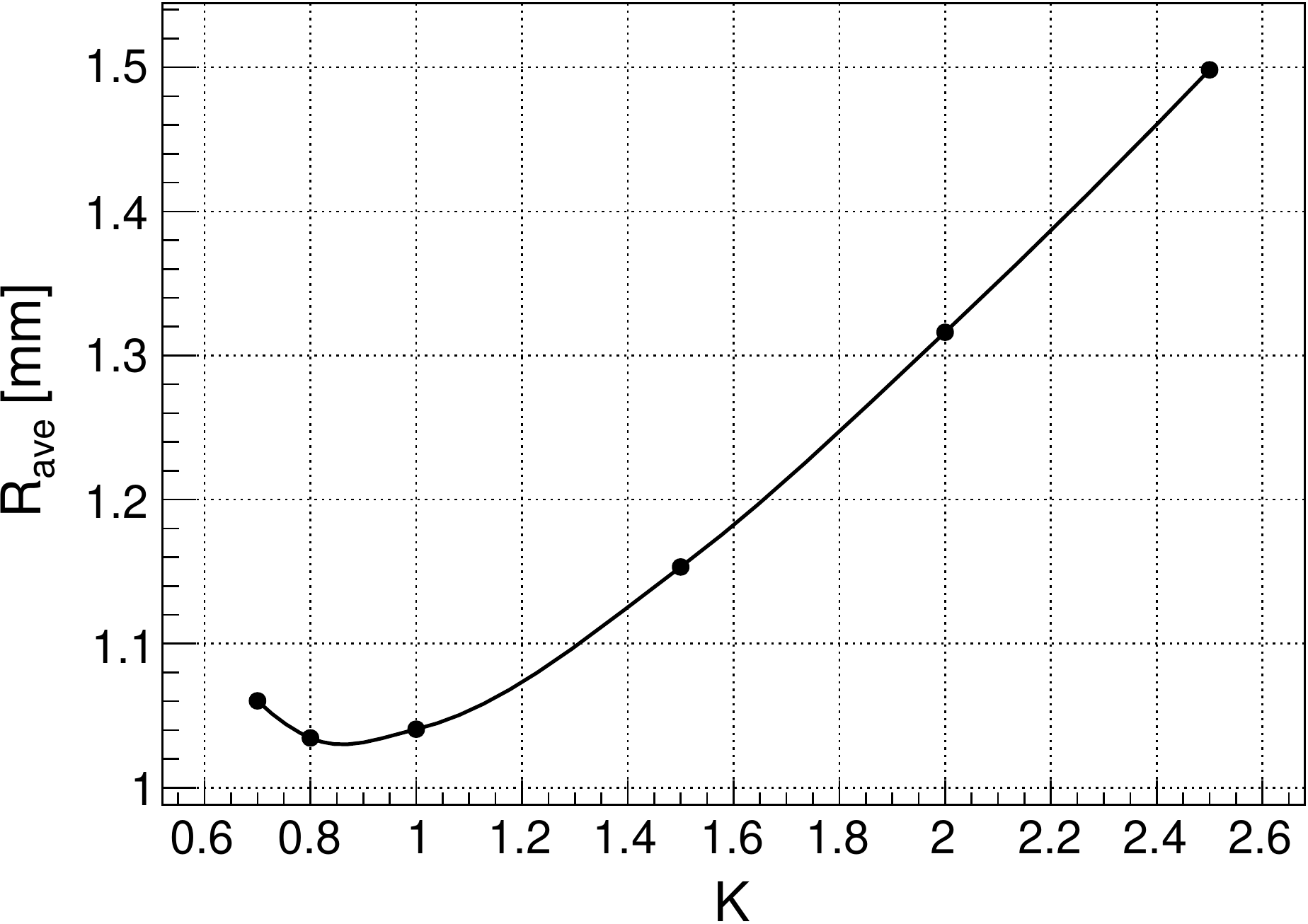}
  \caption{Average power (left) and radius (right) of photon beam on the target.}
  \label{fPhPowerRadius-vs-K-woC}
\end{figure}

Table\,\ref{tNmodules-vs-K-wo} summarizes the required number of active undulator modules and the e$^+$ yield. Figure\,\ref{fPhPowerRadius-vs-K-woC} shows the average photon beam power of the source. The increase of required photon power for high $K$ undulators is connected with a higher spot size of photon beam (Fig.\,\ref{fPhPowerRadius-vs-K-woC}, right plot) and a higher energy of photons resulting in a higher energy spread of the positron beam. This  makes the e$^+$ capture more difficult.

\begin{table}[htb]
  \caption{Required number of active undulator modules and e$^+$ yield vs $K$ for a source without collimator.}
  \label{tNmodules-vs-K-wo}
\renewcommand{\arraystretch}{1.2}
\centering
\begin{tabular}{|l|c|c|} \hline
$K$ & \# Modules & e$^+$ Yield [e$^+$/e$^-$] \\ \hline \hline
0.7 & 21 & 1.564 \\ \hline
0.8 & 16 & 1.500 \\ \hline
1.0 & 11 & 1.521 \\ \hline
1.5 & 6  & 1.586 \\ \hline
2.0 & 4  & 1.655 \\ \hline
2.5 & 3  & 1.688 \\ \hline
\end{tabular}
\end{table}

As it could be seen in Fig.\,\ref{fPol-vs-K-woC}, the maximal polarization of a source without photon collimator is 25\% for 231 meter undulator with $K = 0.7$.
One possibility to get the polarization above 25\% is a further reduction of the undulator field. In this case the undulator has to be longer then 231\,meters. If such elongation of the undulator is not possible or not desired, the capture system must be improved to increase the  polarization.

\begin{figure}[htb]
  \centering
  \includegraphics*[width=85mm]{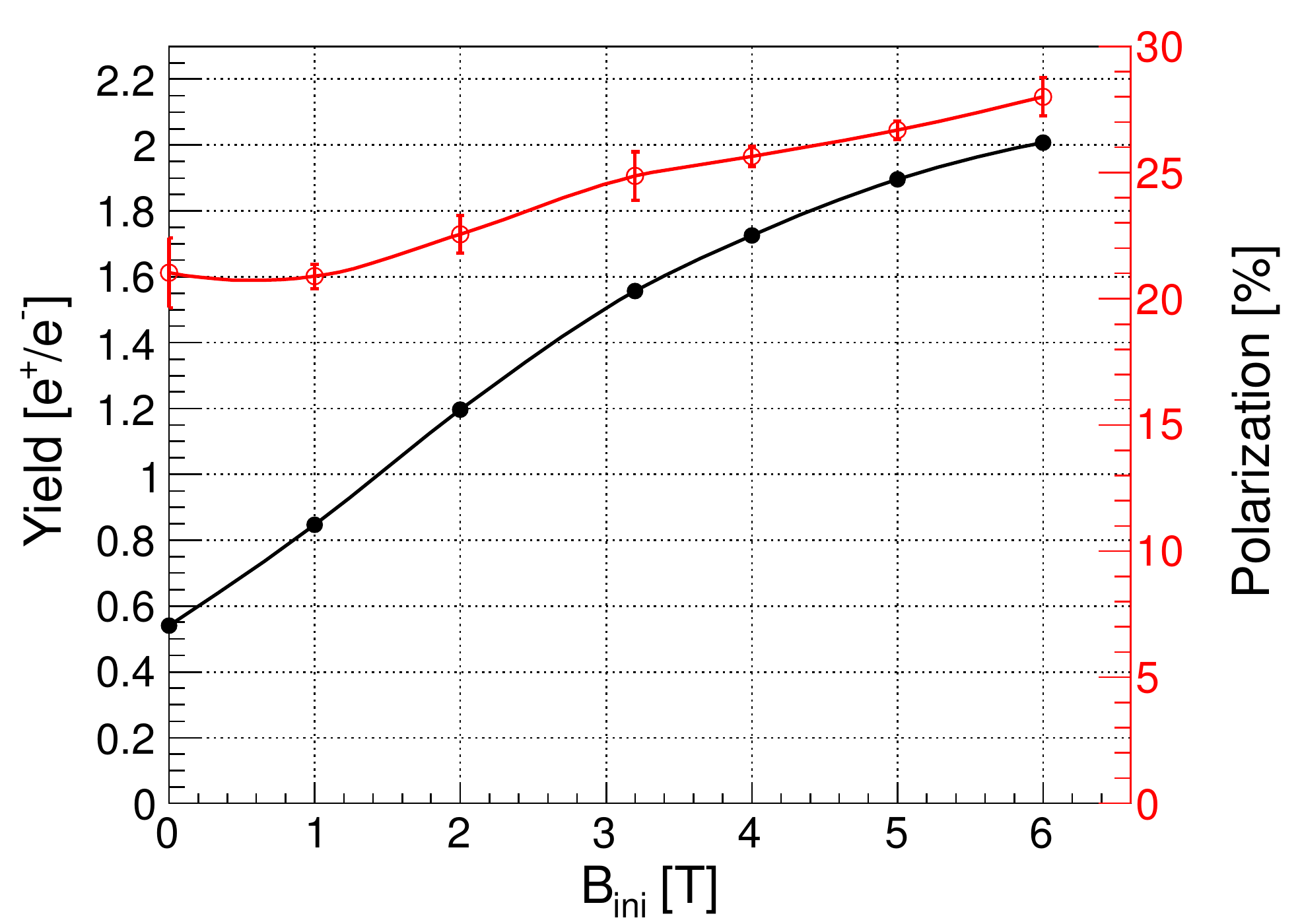}
  \caption{Positron yield and polarization of a source with 231\,m undulator and a flux concentrator with different peak fields at the entrance of FC ($B_{ini}$). FC length is 12\,cm, FC taper parameter is 0.035 mm$^{-1}$.}
  \label{fYandPol-vs-Bfc-woC}
\end{figure}

For instance, a flux concentrator with higher field improves both yield and polarization. Figure\,\ref{fYandPol-vs-Bfc-woC} shows the dependencies of positron yield and polarization on the peak field of a 12\,cm long FC with a taper parameter of 0.035 mm$^{-1}$.

The efficiency of positron generation is growing fast for stronger undulator fields. Figure\,\ref{fPol-vs-K-woC} shows that even for the moderate 3.2 T peak field of FC, the source needs 121 meters undulator with strength $K = 1$; such choice of source parameters results in 20\% positron polarization. Only three undulator modules (33 meters of total active undulator length) are needed to get the required intensity of the positron beam at $K = 2.5$. Hence, there is a big reserve in undulator length in case of using a high $K$ undulator. One possible way to increase the polarization is applying photon collimator~\cite{collimator}. The absorption of photons in the collimator and the reduction of the $e^+$ yield can be compensated by lengthening the undulator.

\begin{figure}[htb]
  \centering
  \includegraphics*[width=80mm]{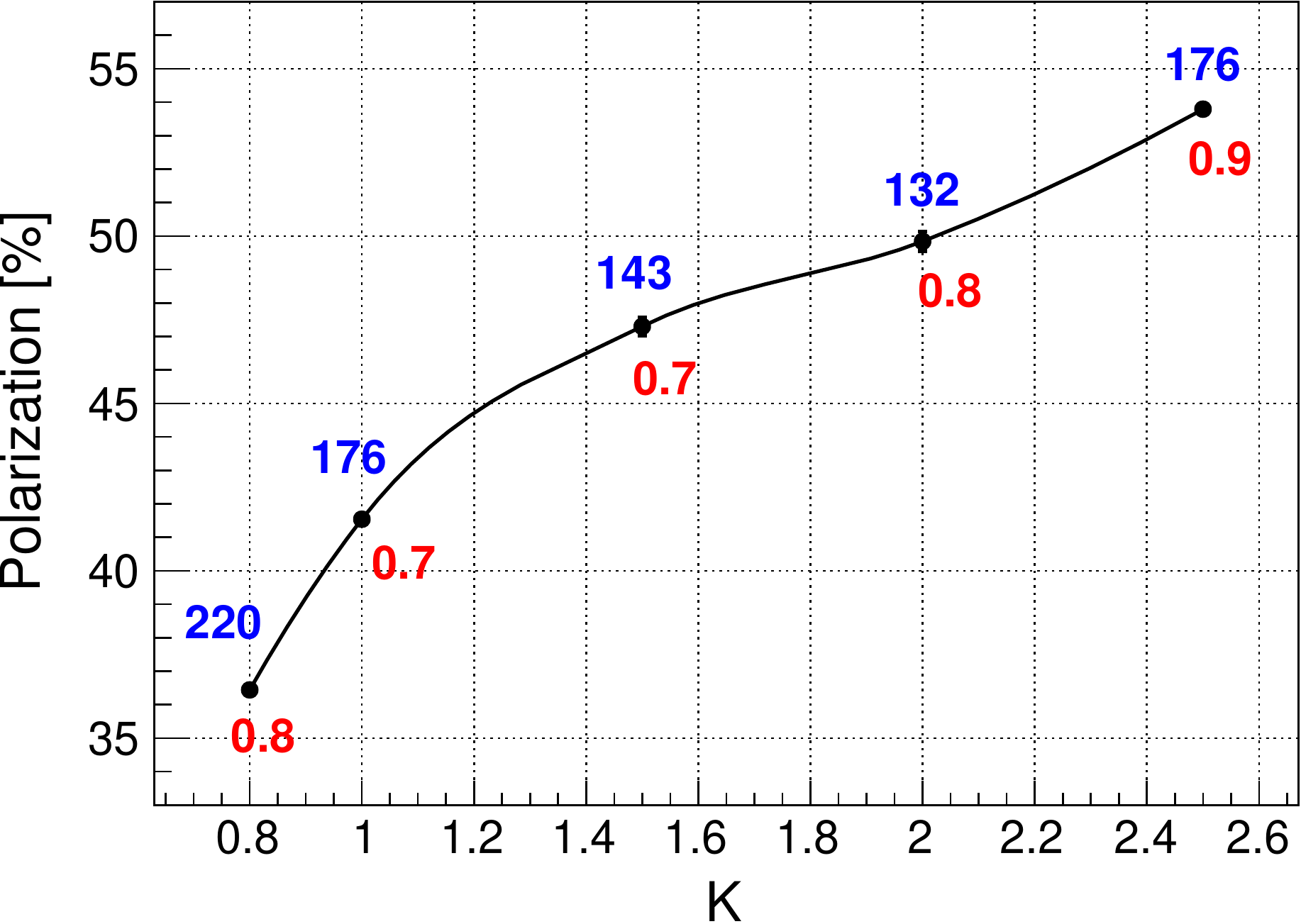}
  \caption{Positron polarization vs $K$ for a source with collimator. The red numbers indicate the aperture radius of collimator and the blue numbers show the required undulator length.}
  \label{fPol-vs-K-withC}
\end{figure}

Figure\,\ref{fPol-vs-K-withC} summarizes the dependence of maximal achievable positron polarization on different undulator $K$ values. In addition to the e$^+$ polarization, also the aperture sizes of photon collimators and the required (for 1.5 e$^+$/e$^-$) undulator lengths are shown as red and blue numbers respectively. 54\% polarization can be reached with a photon collimator having 0.9\,mm aperture radius and $K = 2.5$. The source with $K = 1$ requires a lower collimator aperture ($r = 0.7$\,mm) and the e$^+$ polarization is about 41\%.

\section{Deposited Energy, Temperature Rise and Thermal Stess in Target}

\begin{figure}[htb]
  \centering
  \includegraphics*[width=65mm]{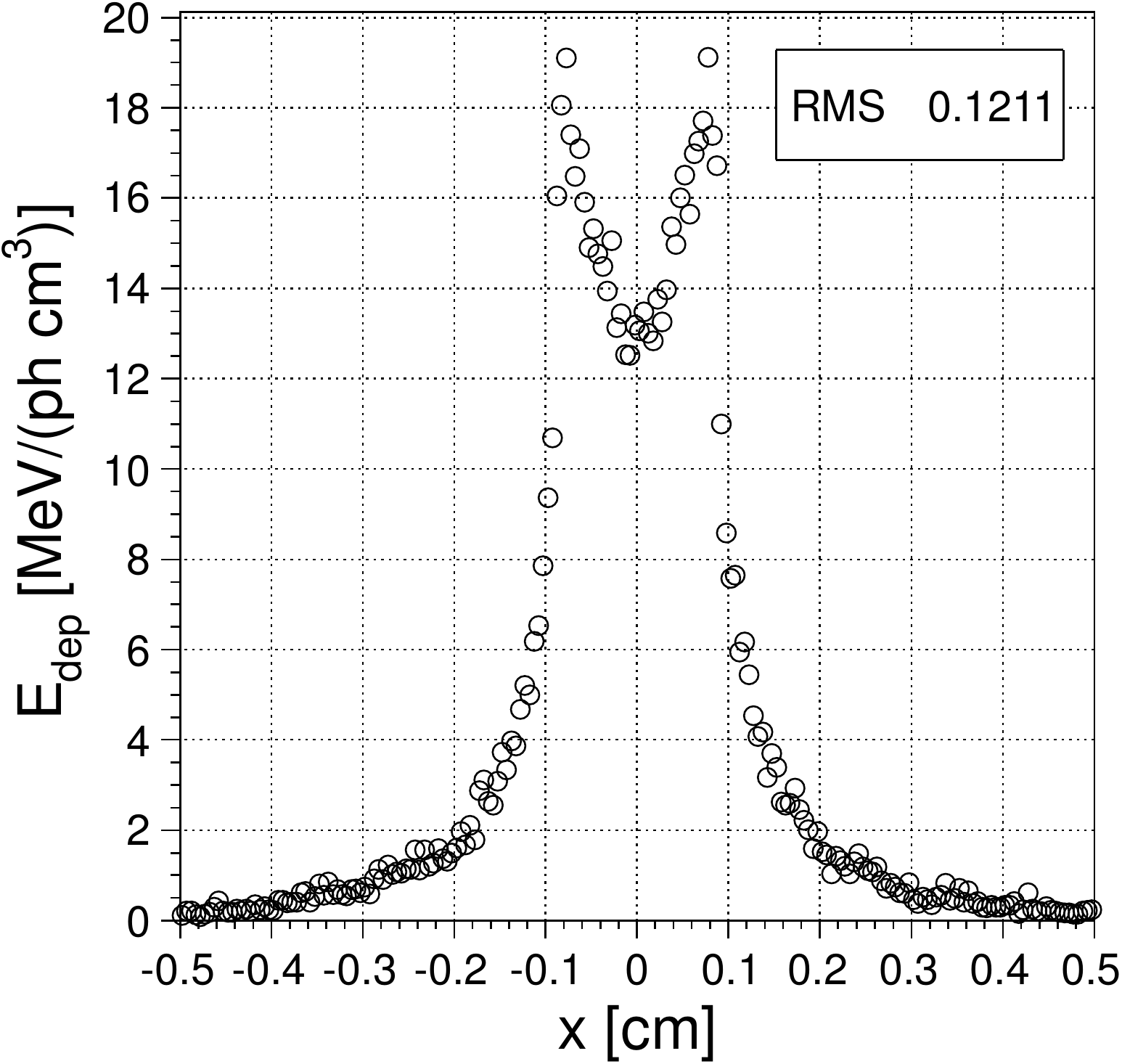}$~~~~~~~$
  \includegraphics*[width=65mm]{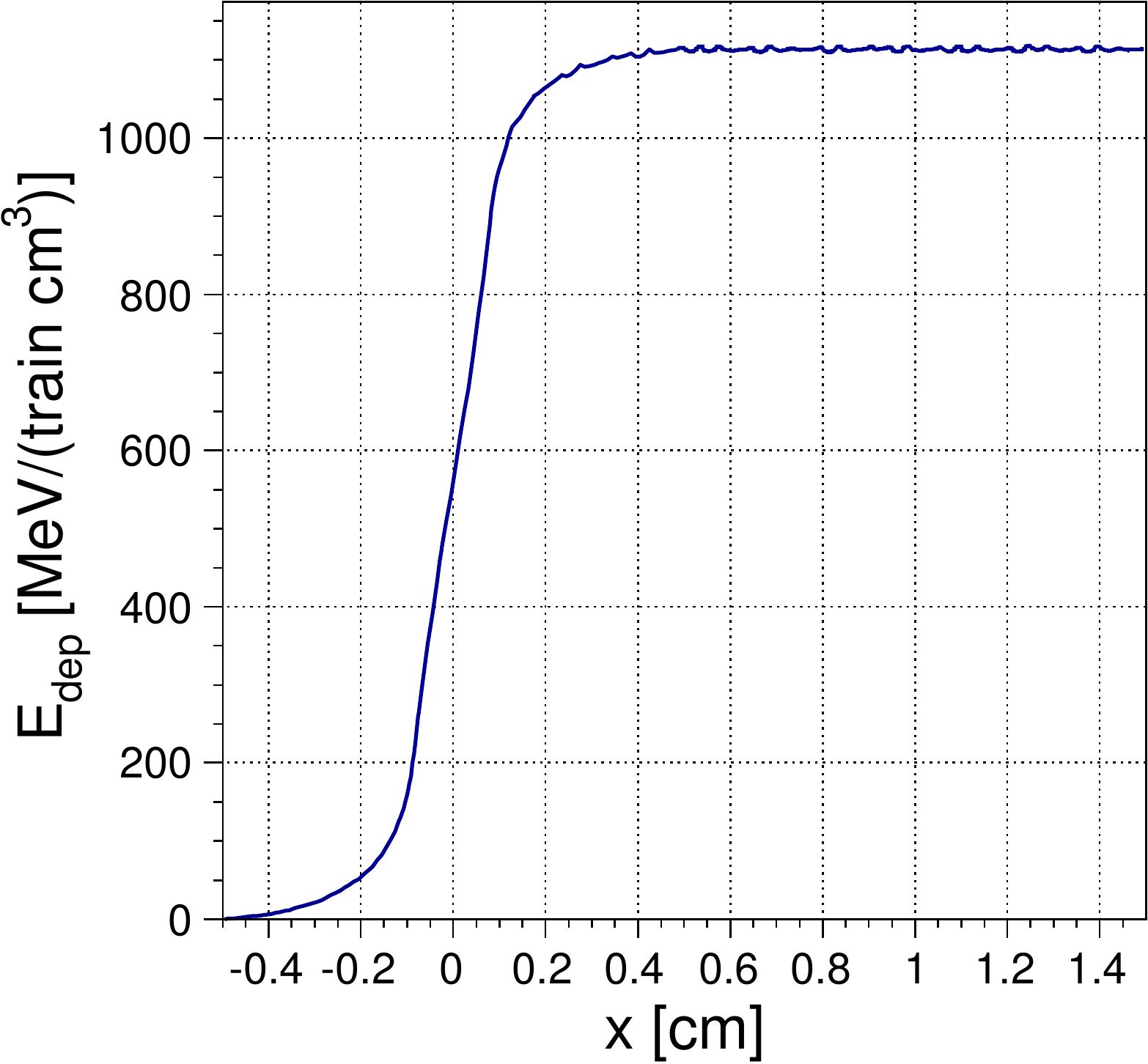}
  \caption{Deposited energy in the target after an one bunch (left) and after one pulse (right).
   Rotation of the target has been modelled as a motion in $x$-direction and the speed $v = 100~$m/s. The undulator has a period $\lambda = 4.3$ cm and $K = 2.5$. The collimator has an aperture $R_c = 0.9~$mm.}
  \label{fEdep-vs-X-K25withC}
\end{figure}

To get 54\% polarization while keeping the yield at 1.5 e$^+$/e$^-$ for a source with $K = 2.5$, the undulator length has to be increased from 33 meters (Fig.\,\ref{fPol-vs-K-woC}) to 176 meters (Fig.\,\ref{fPol-vs-K-withC}). In spite of the major part of photon power will be absorbed in collimator, and the peak energy deposition in the target will be increased significantly. Therefore, the heat load and the thermal stress  induced in the target have to be studied thoroughly.
First, the energy deposition in the target has been calculated in FLUKA for a single bunch (see left plot in Fig.\,\ref{fEdep-vs-X-K25withC}). In this figure, the density of deposited energy is shown as a function of the transverse (to the beam direction) coordinate $x$. In the second step, the target rotation with 100\,m/s tangential speed has been taken into account. To simplify our model, this rotation has been simulated by the motion in $x$-direction only: the distribution shown in Fig.\,\ref{fEdep-vs-X-K25withC} (left) has been shifted in $x$-direction after every bunch. The right plot in Fig.\,\ref{fEdep-vs-X-K25withC} shows the resulting profile of deposited energy along $x$-axis for 366\,ns bunch separation. The energy density in the moving target saturates after few hundred bunches at the level of about 1.2\,GeV per cm$^3$. The ratio of maxima in right and left plots shown in Fig.\,\ref{fEdep-vs-X-K25withC} defines the ``bunch overlapping factor'' for the  rotated target.

\begin{figure}[htb]
  \centering
  \includegraphics*[width=78mm]{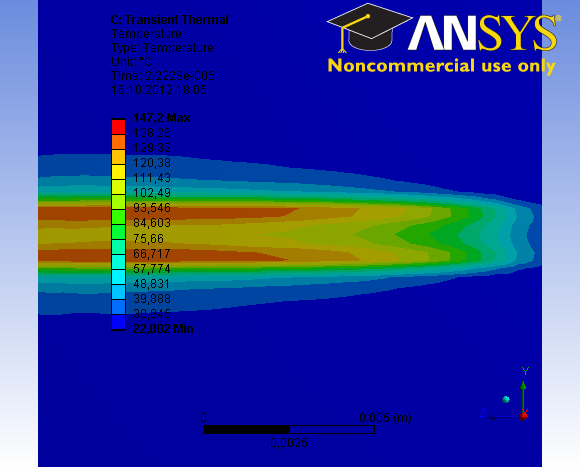}$~$
  \includegraphics*[width=78mm]{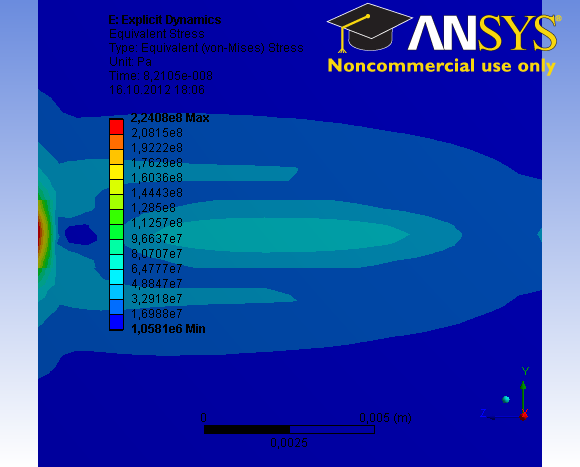}
  \caption{Maximal temperature after first pulse (left) and equivalent von-Mises stress in the rotated target shortly after the pulse has passed the target (right). Undulator period $\lambda = 4.3$ cm, $K = 2.5$; collimator aperture $R_c = 0.9$ mm; target rotation speed $v = 100~$m/s.}
  \label{fTempStress-K25withCrot}
\end{figure}

The three dimensional spatial energy distribution  deposited by one bunch scaled by the number of photons per bunch and multiplied by the bunch overlapping factor has been imported in ANSYS \cite{ANSYS}. The temperature map after a bunch train is shown in Fig.\,\ref{fTempStress-K25withCrot} (left). The maximal rise of temperature after one pulse is about 125$^{\circ}$C. 

The fast increase of target temperature induces thermal stress. The stress distribution in the target shortly after the bunch train passed (82 ns delay) is shown in Fig.\,\ref{fTempStress-K25withCrot} (right). The maximal thermal stress in the target is 224 MPa. This stress is about 25\% of tensile yield stress, and  it is about 44\% of the fatigue stress of untouched Ti6Al4V target material (grade 5, annealed) at $10^7$ cycles. The material properties of titanium alloy were taken from matweb.com database \cite{matweb}. Such stress values (without taking into account the stress due to centrifugal forces of rotating wheel and without accumulating/superposition effects of multiple pulses) can be considered as safe. The region with highest stress is located on the beam axis and close to the back side of target. 

\section{Summary}

The simulations of a positron source with a helical undulator, 4.3\,cm period, using a 500\,GeV electron beam show that a positron beam with 25\% polarization can be generated without photon collimator; only the magnetic field of the undulator has to be reduced ($K = 0.7$). The required undulator length is 231\,m.
The polarization can be increased up to 54\% by applying an undulator with  $K = 2.5$ and a collimator with 0.9\,mm aperture. However, the energy deposited in target and the induced stress are high. So, the maximal thermal stress in the target is increased up to 224 MPa shortly after the photon pulse left the target. It does not destroy the target. To be sure that the target withstands the heat load and mechanical stress during a long time source operation, the model used in ANSYS simulations has to be extended: the centrifugal forces of rotating wheel has to be added and the accumulating/superposition effects of multiple pulses have to be studied too. In addition, a method has to be found to evaluate the fatigue stress and the consequences for the target and collimator  material.

\end{document}